\documentclass[a4paper, 10 pt, conference]{ieeeconf}

\IEEEoverridecommandlockouts

\usepackage{cite}
\usepackage{amsmath,amssymb,amsfonts}
\usepackage{algorithmic}
\usepackage{graphicx}
\usepackage{textcomp}
\usepackage{xcolor}

\usepackage{graphicx}
\usepackage{amsmath}
\usepackage{mathtools}
\usepackage{caption}
\usepackage{multirow}
\usepackage{subcaption}
\usepackage{textcomp}

\usepackage{mdframed}
\usepackage{tabularx}

\newcommand{\RNum}[1]{\uppercase\expandafter{\romannumeral #1\relax}}

\title{\LARGE \bf
Model Checking for Decision Making System of Long Endurance Unmanned Surface Vehicle
}

\author{Hanlin Niu, Ze Ji, Al Savvaris, Antonios Tsourdos, and Joaquin Carrasco
\thanks{*This work was partially supported by EPSRC project No.EP/S03286X/1 and EPSRC RAIN project No. EP/R026084/1. \textit{(Corresponding author: Hanlin Niu)}}
\thanks{H. Niu and J. Carrasco are with the Department of Electrical \& Electronic Engineering, The University of Manchester, Manchester, UK.
        {\{\tt\small hanlin.niu@manchester.ac.uk}\}}%
\thanks{Z. Ji is with the School of Engineering, Cardiff University, Cardiff, UK. 
        {\{\tt\small jiz1@cardiff.ac.uk}\}}
\thanks{A. Savvaris and A. Tsourdos are with the School of Aerospace, Transport and Manufacturing, Cranfield University, Cranfield, UK. 
       {\{\tt\small a.tsourdos@cranfield.ac.uk}\}}
}

\begin{document}

\maketitle
\thispagestyle{empty}
\pagestyle{empty}

\begin{abstract}
This work aims to develop a model checking method to verify the decision making system of Unmanned Surface Vehicle (USV) in a long range surveillance mission. The scenario in this work was captured from a long endurance USV surveillance mission using C-Enduro\textregistered, an USV manufactured by ASV Ltd. The C-Enduro USV may encounter multiple non-deterministic and concurrent problems including lost communication signals, collision risk and malfunction. The vehicle is designed to utilise multiple energy sources from solar panel, wind turbine and diesel generator. The energy state can be affected by the solar irradiance condition, wind condition, states of the diesel generator, sea current condition and states of the USV. In this research, the states and the interactive relations between environmental uncertainties, sensors, USV energy system, USV and Ground Control Station (GCS) decision making systems are abstracted and modelled successfully using Kripke models. The desirable properties to be verified are expressed using temporal logic statement and finally the safety properties and the long endurance properties are verified using the model checker MCMAS, a model checker for multi-agent systems. The verification results are analyzed and show the feasibility of applying model checking method to retrospect the desirable property of the USV decision making system. This method could assist researcher to identify potential design error of decision making system in advance.
\end{abstract}

\section{Introduction}

Unmanned Surface Vehicles can be defined as unmanned vehicles, which execute missions in a variety of hydro environments with least human operation. The guidance, navigation and control systems (GNC) of USV \cite{ren2020finite} allow the marine vessels to follow predefined paths and avoid hazards autonomously \cite{niu2016efficientplanning} \cite{zhu2020adaptive} \cite{zhu2019optimized} \cite{zhu2018identification}, relieving the operators from the heavy and tedious manual operations. Further development of USVs are expected to produce tremendous benefits, such as lower operation costs, improved energy efficiency, personnel safety and security, extended operational reliability and precision, as well as increased flexibility in complex environments, including so called dirty, dull, harsh, and dangerous missions \cite{bertram2008unmanned} \cite{niu2019voronoi}. To improve the operating endurance, USVs  powered by multiple sources of energy are developed, utilising solar energy, wave energy or wind energy \cite{makhsoos2019evaluation} \cite{ren2021active}.

Because of the critical nature of USV decision making system in long endurance missions, it is important to ensure the correctness of the decision making system. Verification is the process of verifying the correctness of the system by checking against the specifications \cite{sirigineedi2010kripke} \cite{webster2014generating} \cite{ezekiel2011verifying} \cite{molnar2009system}. Typical verification processes include simulation, testing, deductive verification, and model checking \cite{Clarke:2001:MC:778522.778533}. Simulation is implemented using the abstract model of the system and testing is performed on the real system. The simulation and testing are a cost effective way to identify bugs. However, it is not practically possible to check all cases exhaustively. Deductive verification is proof-based and it is well recognised by computer scientists. However, it is time-consuming and can only be performed by the experts in logics and mathematics. Model checking is a kind of formal verification methods that are usually used for exhaustive system analyses automatically to check whether the model of the system satisfies the desirable properties.

Model checking has been implemented in the verification of autonomous systems \cite{choi2015verification} \cite{barbier2019validation} \cite{clarke2018handbook} . NASA has developed model checking techniques for multiple rovers or satellites \cite{brat2005challenges} \cite{pecheur2000verification}. In the work of \cite{quottrup2004multi}, timed automata have been applied to model multiple robotic systems, where the properties are expressed in CTL (Computational Tree Logic) and finally verified by the Uppaal model checker. The Kripke model of a single UAV (Unmanned Aerial Vehicle) performing a search mission was modelled in \cite{sirigineedi2009towards} and the properties expressed in CTL have been verified using Symbolic Model Checker(SMV). Subsequently, the scenario was extended to a multiple UAV searching scenario in \cite{sirigineedi2011kripke}. A multiple UAV system monitoring road networks was modelled and verified in \cite{sirigineedi2010kripke}. A group of robots operate with minimal communication with no priori knowledge of the environment has been modelled using the Kripke model \cite{jeyaraman2006kripke}. The desirable properties of co-operation were expressed using Linear Temporal Logic(LTL) and the properties were finally verified using SPIN. The integration of model checking methods with UAV mission planning systems was proposed in \cite{humphrey2013model} and it enables the autonomy to make decisions by human intent and provides better feedback to the human when problems arise. Another USV mission plan verification for a VIP escort mission was presented in \cite{humphrey2012model} that, in this scenario, multiple UAVs should monitor and navigate a ground-based VIP vehicle to follow a road network. The model was built using PROcess MEta LAnguage(PROMELA), and the properties were expressed using LTL and verified using SPIN, which is a multi-threaded model checker.

The main contribution of this paper is the implementation of the model checking method on the verification of the real USV system. The mission considered was captured from the C-Enduro USV \cite{ASV2017} surveillance case, which was funded by the UK government-backed Small Business Research Initiative (SBRI). This paper presents the process of modelling the behaviours and the complex reactive relations among multiple environmental factors, the corresponding sensors, the energy system, the USV and the GCS decision making systems. The complex environment is discretised and abstracted using the Kripke model, which is a formal and intuitive model in the form of a directed graph \cite{Clarke:2001:MC:778522.778533} \cite{huth2004logic}. The behaviours of the USV are also classified based on the energy required. The desirable properties of the decision making system are expressed precisely using \textit{CTL}. Finally, the feasibility of using the model checker MCMAS to verify the safety property and the long-endurance/energy-saving property of the USV decision making system is demonstrated. The remainder of this paper has the following structure: The USV mission scenario is presented in section \ref{sec:scenario}. Section \ref{sec:kripke} introduces the Kripke models of the environmental factors and the autonomous systems. In section \ref{sec:MCMAS}, the desirable properties are expressed using \textit{CTL} and verified using MCMAS. Finally, the conclusion and future work are given in section \ref{sec:conclusion}.

\section{Mission Scenario} \label{sec:scenario}

\begin{figure}[!ht]
  \begin{center}
    \includegraphics[width=3.4in]{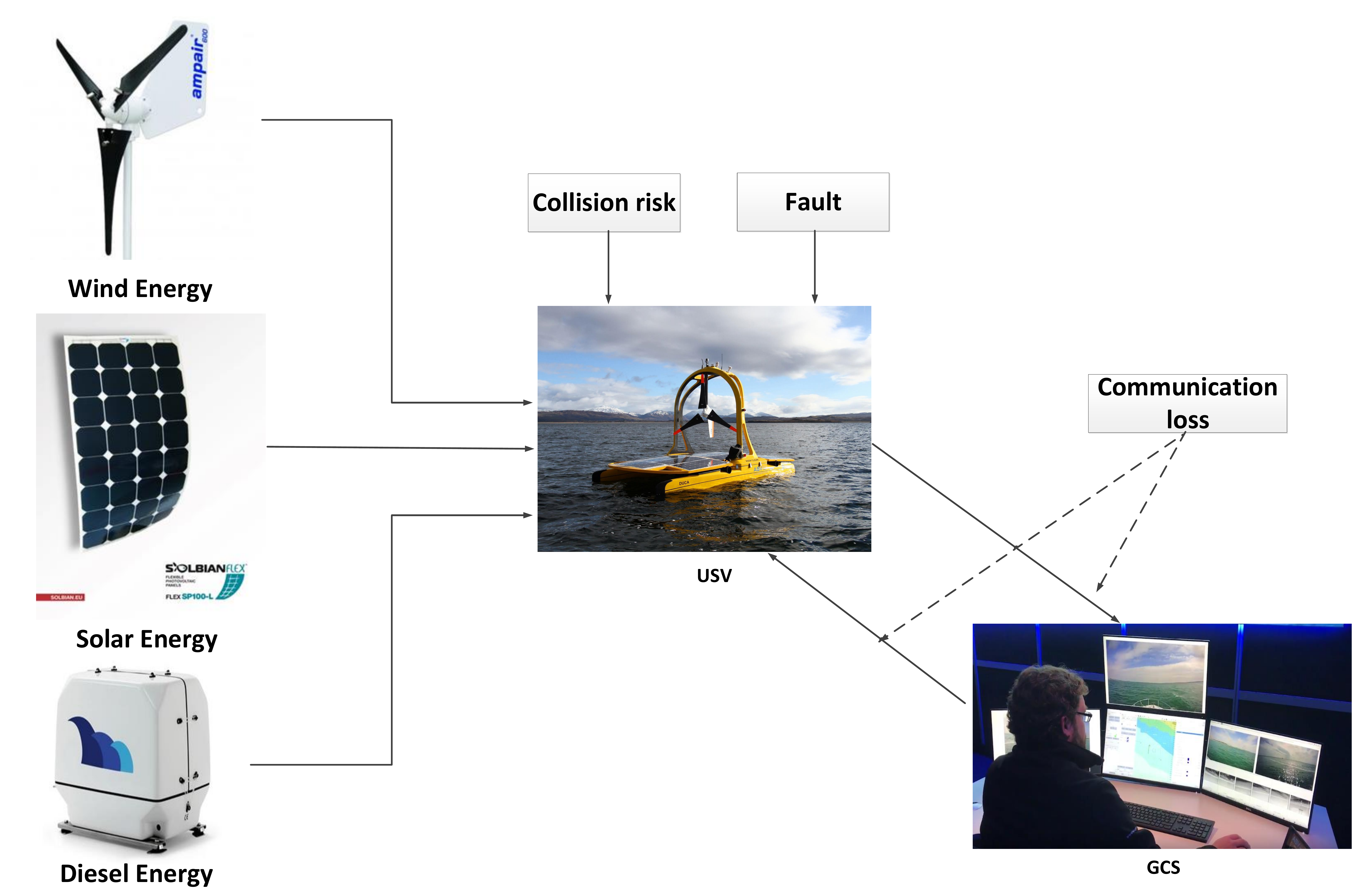}
    \caption{The USV mission with multiple energy sources \cite{ASV2017}}
    \label{fcenario}
  \end{center}
\end{figure}

The mission scenario and decision making system considered in this paper were captured from the long endurance USV, C-Enduro, as depicted in Fig.~\ref{fcenario}. USV is commanded by GCS to execute a long range surveillance mission by following a list of waypoints, which are generated by an energy efficient path planning algorithm \cite{niu2018energy} \cite{niu2020energy} of the GCS. While USV is following the path, it sends images to the GCS for analysis. USV may encounter problems including communication signal loss, collision risk and malfunction. The decision making systems of the USV and GCS are required to ensure safety and also maximise the utilisation of natural energy for long endurance operations by adaptive USV behaviours.

\section{Kripke Modelling} \label{sec:kripke}
During the long range marine mission, to model the complex environment in terms of communication, traffic, malfunction and energy, we present an approach to discretise and abstract the environments using non-energy-related and energy related models, which helps in reducing the state space. Finally, the Kripke models of the the USV decision making system and the GCS decision making system are presented.

\subsection{Kripke models of non-energy-related environmental uncertainties and the corresponding USV sensors}

Non-energy-related environmental factors interacting with the USV system include communication signal and traffic information. Malfunction of USV is also modelled as an environmental factor. The states of the communication signals can be identified by using heart-beat messages and we call this signal detection mechanism the communication detector. Traffic information can be detected by AIS (Automatic Identification System) sensors. It is assumed that malfunction can be detected by the USV online and we call this mechanism as fault detector. In this research, we assume the sensors can detect the corresponding environmental factors accurately.

\subsubsection{Kripke models for communication signal and communication detector}

\begin{figure}[!ht]
  \begin{center}
    \includegraphics[width=2in]{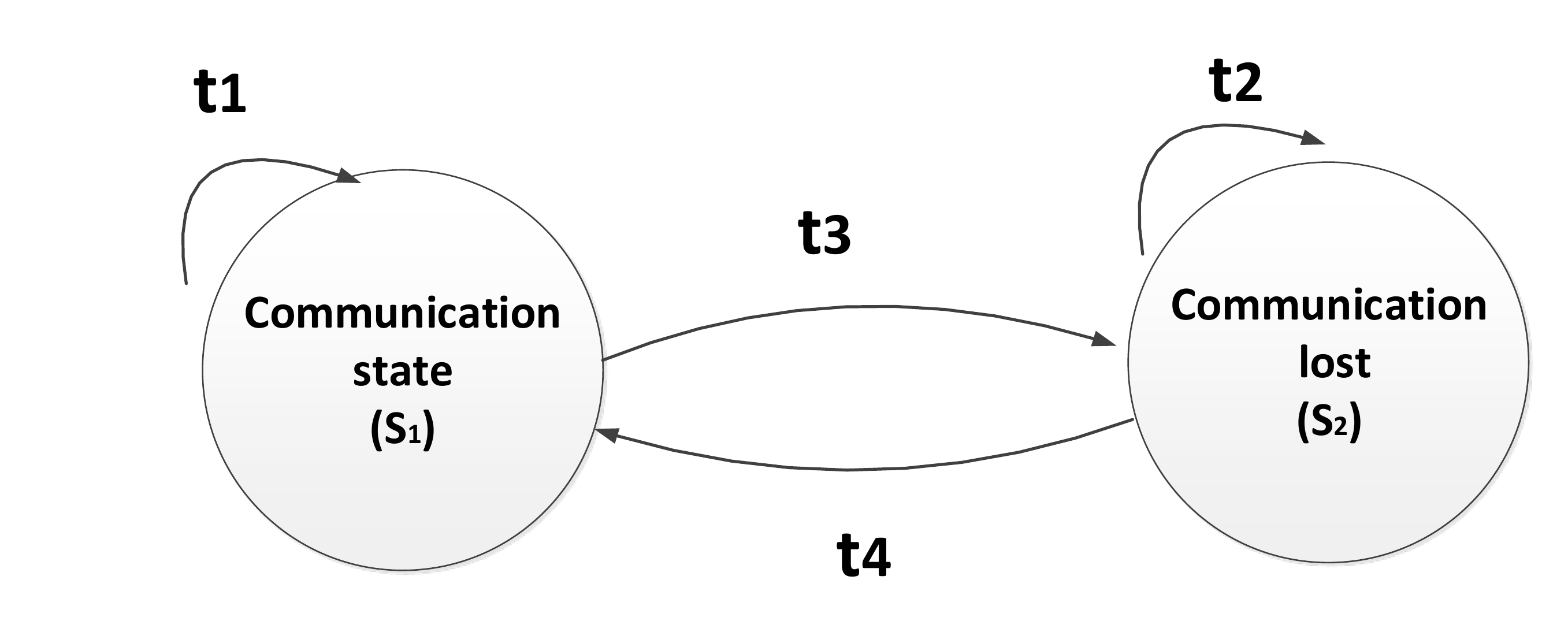}
    \caption{Kripke model for the communication channel}
    \label{f5}
  \end{center}
\end{figure}

In Fig.~\ref{f5}, the behaviours of the communication channel between the USV and the GCS can be defined as two states: \textit{communication state} and \textit{communication lost state}, which are represented by symbol $S$. The communication channel is treated as a non-deterministic system, which means each state at a specific moment may have multiple possible consequential transitions. For example, the \textit{communication state} has two allowed transitions, namely $t_{1}$ and $t_{3}$, which transit the current state to \textit{communication state} or \textit{communication lost state} respectively. Similarly, at the moment of \textit{communication lost state}, it can also have two transitions, namely $t_{2}$ and $t_{4}$, with the next state as either \textit{communication state} or \textit{communication lost state}. This non-deterministic model obeys the real situation that the state at each specific moment may transit to multiple possible states.

The conditions for state transitions are given as follow:
\begin{itemize}

\item $t_{1}, t_{4}$: If the communication is normal.

\item $t_{2}, t_{3}$: If the communication state is lost.

\end{itemize}

The states of the communication channel can be detected by the communication detector. The states of the communication detector are defined as \textit{communication state detected} and \textit{communication lost state detected}. The transitions of detector will take place with the changes of communication states.

\subsubsection{Kripke model for collision risks and AIS}

Collision risks can be classified into two categories according to COLREGS (International Regulations for Preventing Collisions at Sea) \cite{savvaris2014development}, namely \textit{give-way collision risk} and \textit{stand-on collision risk}. The \textit{give-way collision risk} and \textit{stand-on collision risk} represent the collision scenarios that the USV should give its way to or keep the way to avoid collision with the encountered vessel, respectively. Therefore, the traffic situation can be modelled using three states including \textit{no collision risk}, \textit{give-way collision risk} and \textit{stand-on collision risk}. The traffic situation is also a non-deterministic system. The traffic information can be detected by AIS sensors. The states of the AIS are defined as \textit{no collision risk detected}, \textit{give-way collision risk detected} and \textit{stand-on collision risk detected}. The state transitions of the AIS are triggered by the transitions of traffic information.

\subsubsection{Kripke model for fault event and fault detector}
The states of fault events include \textit{severe fault}, \textit{fault} and \textit{non-fault}. During the long range mission, the USV may encounter malfunction but can still have the collision avoidance capabilities. Therefore, a distinction between \textit{severe fault} and \textit{fault} is required to improve the safety of the USV. The state \textit{severe fault} is defined to represent that the USV cannot operate anymore and it will go to \textit{standby} immediately. The state \textit{fault} event is emitted when the USV cannot execute the path following command but still execute the collision avoidance command. In this situation, the USV will remain at the \textit{station keeping} state. \textit{Non-fault} means the USV is in normal operation state. The fault event is also treated as a non-deterministic system. The states of fault detector include \textit{Severe Fault detected}, \textit{Fault detected} and \textit{Non-fault detected}. State transitions happen with corresponding changes of the fault events.

\subsection{Kripke models of energy-related environmental uncertainties and the USV energy system}

Energy-related environmental factors have impact on the energy generation and include the solar irradiance and the wind conditions in this work. Instead of modelling these two factors separately, we name these two factors' model as the energy generation condition model by referring to the total influence of them on the energy system. Since the vehicle in this work is powered by solar panel, wind turbine and diesel generator using the natural resources or the fuel, we name these three equipments as the energy generation module. The environmental factors that have the largest impact on the energy consumption is the sea current that the USV encounters. We modelled the sea current condition as energy consumption condition model. The states of the energy generation module and energy consumption module will contribute to the transitions of the battery level. Therefore, the whole energy model can be divided into five sub-models: energy generation condition model, energy generation module model, energy consumption condition model, energy consumption module model and battery model.

The energy generation module can be affected by the energy generation conditions and battery level. The diesel generator will be turned on or off subject to the status of battery level. The states of the energy consumption module model can be affected by the energy consumption condition (sea current condition) and USV behaviours. Therefore it is necessary to classify the USV behaviours based on the energy consumption characteristics. The states of the energy generation module and the energy consumption module will contribute to the transitions of the battery level. Finally, the states of the energy generation module, the energy consumption module and the battery will affect the decision making system of the USV.

The energy model used in this scenario is proposed by referring to the energy consumption specifications and energy generation specifications of the C-Enduro USV. We abstracted and discretised the energy consumption model, energy generation model and battery model by using integers to represent the amount of the energy, which helps in reducing the computational state space.

\subsubsection{Energy generation}

\begin{figure}[!ht]
  \begin{center}
    \includegraphics[width=3in]{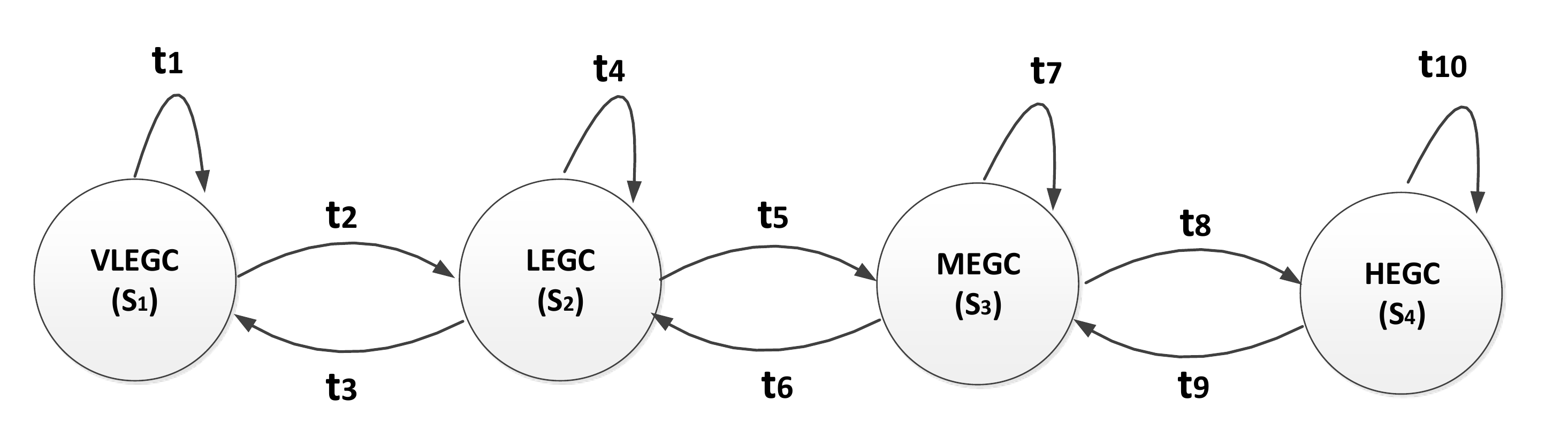}
    \caption{Kripke model for energy generation condition}
    \label{fenergy-generation-condition}
  \end{center}
\end{figure}

Four energy generation conditions are modelled: \textit{Very Low Energy Generation Condition (VLEGC), Low Energy Generation Condition (LEGC), Medium Energy Generation Condition (MEGC)} and \textit{High Energy Generation Condition (HEGC)}. The energy generation condition can change from one state to its neighbour state randomly, as shown in Fig.~\ref{fenergy-generation-condition}. Correspondingly, there are four states with the energy generation module: \textit{Very Low Energy Generation (VLEG)(+0), Low Energy Generation (LEG)(+1), Medium Energy Generation (MEG)(+2)} and \textit{High Energy Generation (HEG)(+3)}. These four states of energy generation module correspond to the amount of energy generation that will be added to the battery level. For instance, when the energy generation condition is in \textit{VLEGC} state, the state of the energy generation will be \textit{VLEG} correspondingly and the energy added to the battery will be 0. Note that when the diesel generator is on, the energy generation state is always \textit{HEG}. This is defined according to the specification of the C-Enduro diesel generator.

\subsubsection{Energy consumption}

The energy consumption state is modelled by discretising the energy consumption condition state (the sea current state) and classifying the USV state. The states of the environmental energy consumption conditions and the behaviours of the USV will determine the states of the energy consumption module. Three environmental energy consumption conditions are modelled: \textit{Low Energy Consumption Condition (LECC), Medium Energy Consumption Condition (MECC)} and \textit{High Energy Consumption Condition (HECC)}. The energy consumption condition can transit from one state to its nearby state randomly. The behaviours of the USV can be classified into three groups according to the amount of the corresponding energy consumption, which includes \textit{Low Energy Consumption Behaviour (LECB)} (\textit{Station Keeping (SK)}), \textit{Medium Energy Consumption Behaviour (MECB)}, (\textit{Path Following (PF), Collision Avoidance (CA)) } and \textit{High Energy Consumption Behaviour (HECB) (Path Following in High Speed (PFH))}. Note that other USV behaviours, including \textit{Standby (SB), Ready (RE), Dispatched (DP), Arrive (AR)}, are treated separately because they consume very little energy and the energy consumption effect will be negligible by the environmental factors. For simplification, we assume that the energy consumption amount of these behaviours is 0.

Various kinds of combinations of the energy consumption conditions and behaviours of the USV will lead to the corresponding state transitions of the energy consumption module, including the amount of the battery level to be subtracted and the states of the energy consumption module. The amount of energy consumption is given as following: \textit{Very Low Energy Consumption (VLEC) (-0), Low Energy Consumption (LEC) (-1), Medium Energy Consumption (MEC) (-2), High Energy Consumption (HEC) (-3)} and \textit{Very High Energy Consumption (VHEC) (-4)}. The relations between the energy consumption condition, the USV behaviour and the energy consumption amount is shown in Table \ref{energy-consumption-table}, which is self-explanatory. For instance, when the USV is in \textit{Low Energy Consumption Behaviour (LECB)} and the energy consumption condition is also \textit{Low Energy Consumption Condition (LECC)}, the consumed energy will be \textit{Very Low Energy Consumption (VLEC)}.

\begin{table}[h]
  \caption{The relations between energy consumption conditions, USV behaviours and energy consumption amount}
  \label{energy-consumption-table}
  \begin{center}
    \begin{tabular}{|c||c||c||c|}
      \hline
      & \textit{LECC} & \textit{MECC} & \textit{HECC} \\
      \hline
      \textit{LECB} & \textit{VLEC} (0) & \textit{LECC} (-1) & \textit{MECC} (-2) \\
      \hline
      \textit{MECB} & \textit{LEC} (-1)& \textit{MEC} (-2)& \textit{HEC} (-3) \\
      \hline
      \textit{HECB} & \textit{MEC} (-2) & \textit{HEC} (-3) & \textit{VHEC} (-4)  \\
      \hline
    \end{tabular}
  \end{center}
\end{table}

\subsubsection{Battery}
The battery level is represented by an integer from 0 to 10. The accumulation of the energy consumption amount and the energy generation amount will be the changing amount of the battery level. For example, if the current state of the battery level is 5, the state of the energy generation module is \textit{Low Energy Generation (LEG, +1)} and the state of the energy consumption module is \textit{Medium Energy Consumption (MEC, -2)}, then the next state of the battery level will be updated to be $(5+1-2 = 4)$.

\subsection{Kripke model for USV}

The behaviours of the USV are defined as follows: \textit{SB, RE, DP, PF, PFH, CA, SK, SFA(Severe Fault), FA(Fault)} and \textit{AR}. In this mission scenario, the battery level is taken into account in the decision making system. When the battery level is 0 and 1, the USV should be in \textit{SB} or \textit{SFA} state and the diesel generator will be triggered to generate power. When the battery level is 2, the USV can be \textit{SK}, \textit{CA}, \textit{SB} or \textit{SFA} state and turn off the diesel generator. When the battery level is above 3, the USV can be in \textit{PF, SK, CA, RE, DP, SB} or \textit{FA} state. When the battery level is above 9 and the energy consumption condition is \textit{LECC} and the energy generation condition is \textit{HEGC}, the USV will choose \textit{PFH} state other than \textit{PF} for maximising the utilisation of natural energy. The transitions and the corresponding conditions are described as follows:

\begin{figure}[!ht]
  \begin{center}
    \includegraphics[width=3.5in]{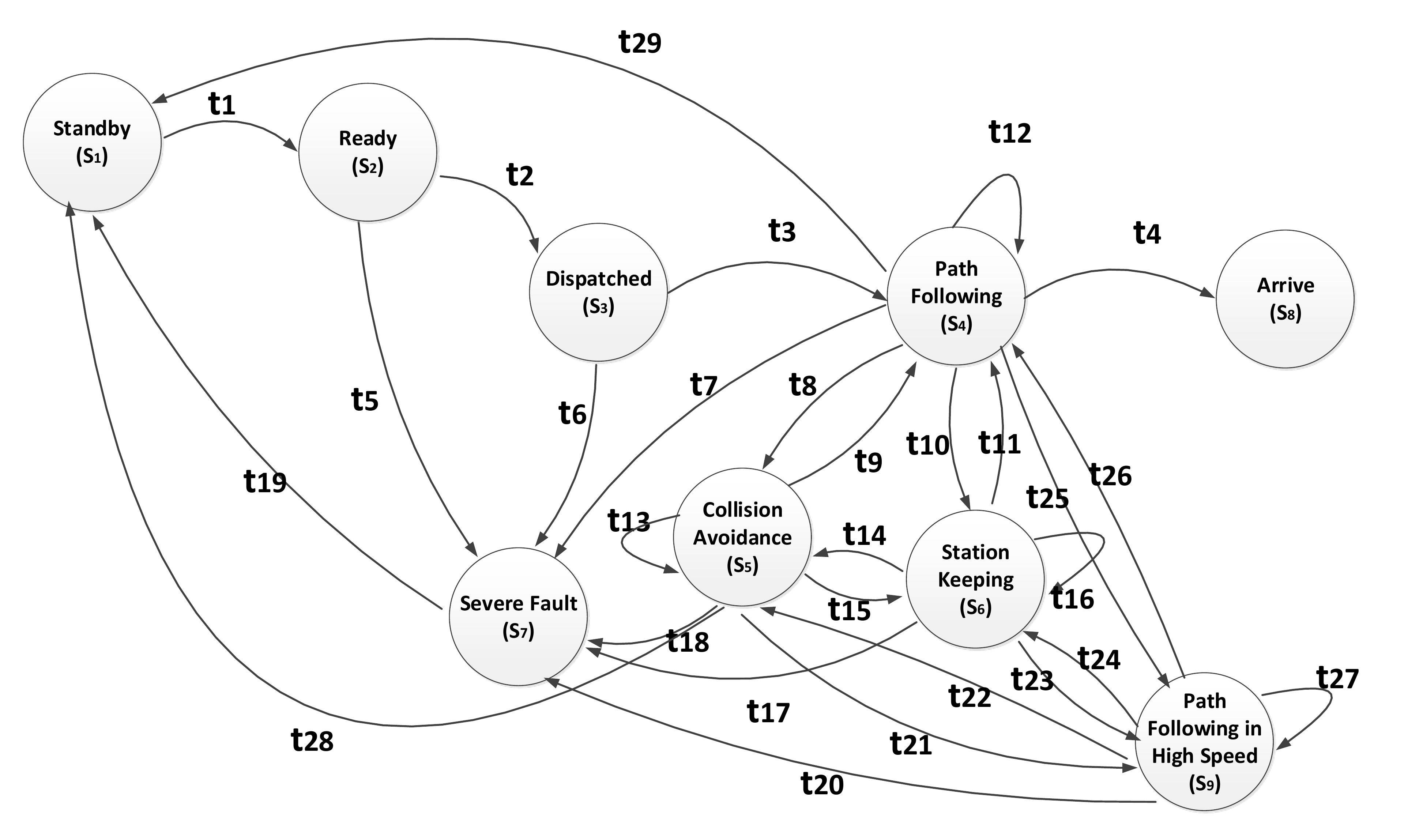}
    \caption{Kripke model for the USV}
    \label{fUSV}
  \end{center}
\end{figure}

\begin{itemize}
\item $t_{1}$: If the USV received the mission from the GCS, no fault is detected and battery level is above 2.

\item $t_{2}$: If the USV received the launching command from GCS and no fault is detected.

\item $t_{3}$: If the USV has been dispatched and no fault is detected.

\item $t_{4}$: If the USV arrived the destination.

\item $t_{12}, t_{9}, t_{11}, t_{26}$: If the USV is in \textit{PF}, \textit{PFH}, \textit{CA} or \textit{SK}, no giving-way collision risk is detected; communication channel is in good status; no fault is detected; and the battery level is above 2.

\item $t_{8}, t_{13}, t_{14}, t_{22}$: If the USV is in \textit{PF}, \textit{PFH}, \textit{CA} or \textit{SK}, a giving-way collision risk is detected; communication channel is in good status; no fault is detected; and the battery level is above 1.

\item $t_{10}, t_{15}, t_{16}, t_{24}$: If the USV is in \textit{PF}, \textit{PFH}, \textit{CA} or \textit{SK}, no giving-way collision risk detected; communication channel is lost; no severe fault is detected; and the battery level is above 1; \textbf{or} fault event is detected and no collision risk is detected; \textbf{or} no fault is detected; the battery level is 2; and no collision risk is detected.

\item $t_{5}, t_{6}, t_{7}, t_{18}, t_{17}, t_{20}$: If the USV has detected severe faults.

\item $t_{19}$: If the USV is in the \textit{SFA} state.

\item $t_{28}, t_{29}$: If the USV battery level is 0 or 1.

\item $t_{25}, t_{23}, t_{21}, t_{27}$: If there is no give-way collision risk; no fault detected; battery is above 8; and the energy generation is higher than energy consumption.

\end{itemize}

\subsection{Kripke model for GCS}
The behaviours of the GCS are defined as follows: \textit{Path Planning (PP), Send Waypoints (SW), Launch Command (LC), Situation Analysis (SiA), Path Re-planning (PR)} and \textit{Send New Waypoint (SN)}. Fig.~\ref{fGCS} shows the Kripke model of the GCS behaviours. The transitions and the corresponding conditions are described as follows:

\begin{figure}[!ht]
  \begin{center}
      \includegraphics[width=3.2in]{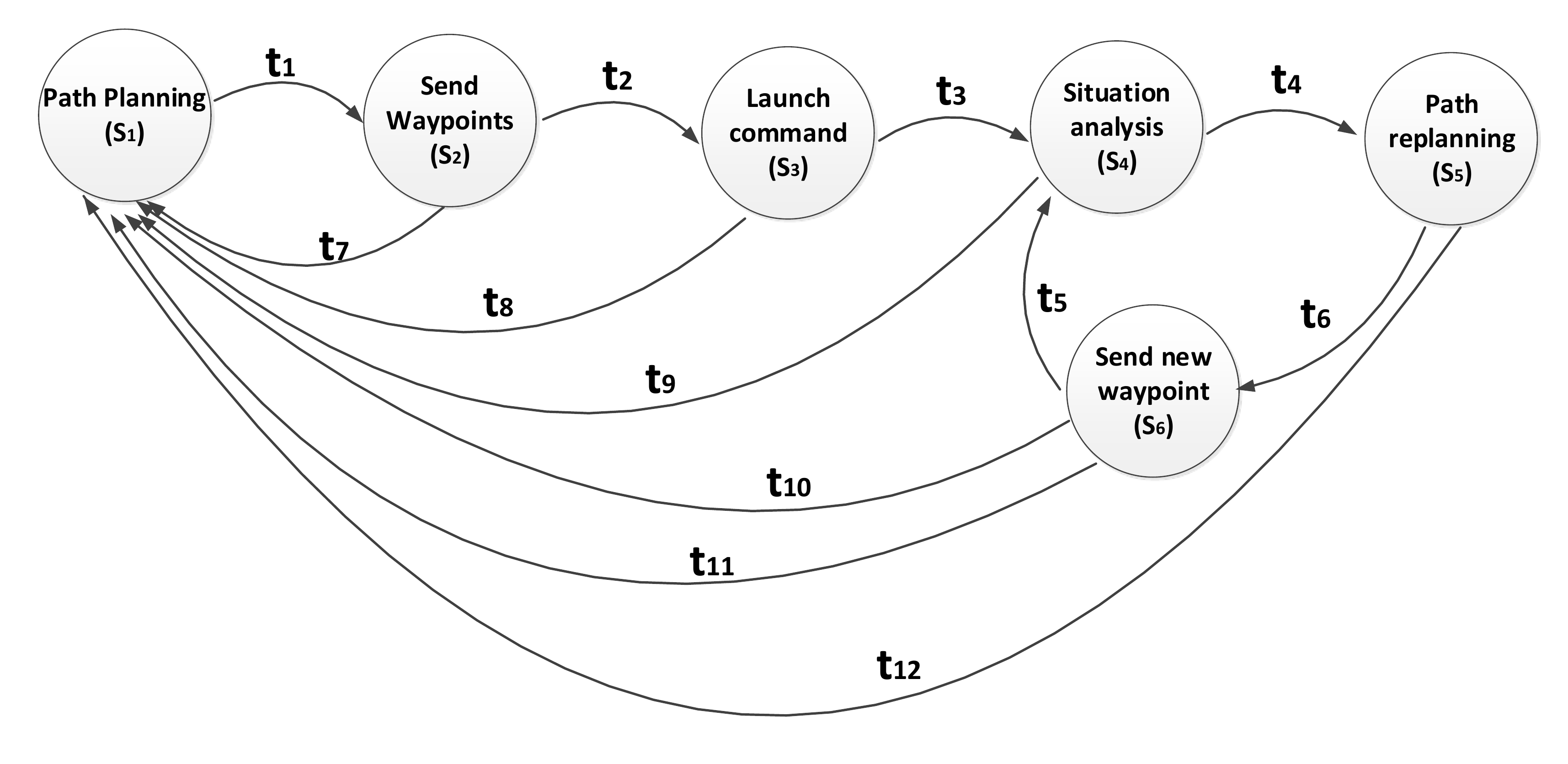}
    \caption{Kripke model for the GCS}
    \label{fGCS}
  \end{center}
\end{figure}

\begin{itemize}
\item $t_{1}$: If the GCS is in \textit{PP} state and the USV is in \textit{SB} state.

\item $t_{2}$: If the GCS is in \textit{SW} and the USV is in the \textit{DP} state.

\item $t_{3}$: If the GCS is in \textit{LC} state and the USV is in the \textit{PF} state.
  
\item $t_{4}$: If the GCS is in the \textit{SiA} state, the USV is in the \textit{SK} state and the communication state gets recovered.

\item $t_{6}$: If the GCS is in the \textit{PR} state.

\item $t_{5}$: If the GCS is in the \textit{SN} state and the USV is in the \textit{PF} mode.

\item $t_{7}, t_{8}, t_{9}, t_{10}, t_{11}, t_{12}$: If the USV has detected fault and the communication status is normal.

\end{itemize}

\section{Model Checking with MCMAS} \label{sec:MCMAS}

The Kripke models are translated into the ISPL code, the modelling language of MCMAS, which is an open-source model checker designed for verification of Multi-Agent Systems. The desirable properties are expressed using $CTL$ formulaes and implemented into the \textit{Evaluation} and \textit{Formulae} part of MCMAS. Finally the properties are verified using MCMAS. The details are presented in the following subsections.

\subsection{MCMAS model}

MCMAS uses its own language ISPL to describe the system model. ISPL has six essential parts including \textit{Environment Agent, Agent, InitStates, Evaluation}, and \textit{Formulae}. In the \textit{Environment Agent} and \textit{Agent}, the possible states, the labelling function and transitions of the Kripke model can be parsed using \textit{state variables}, \textit{actions}, \textit{protocols} and \textit{evolution}. The \textit{InitStates} defines the initial states of all agents. The atomic propositions of the properties to be verified are declared in \textit{Evaluation}. These propositions and \textit{CTL} are used to describe how the behaviours of the system unfold over time. The properties that we want to verify are expressed in the \textit{Formulae} part. The states of the \textit{Agents} are described in \textit{Vars}. Each \textit{Agent} is allowed to perform some \textit{Actions}, which are visible by other \textit{Agents}. The \textit{Actions} correspond to the atomic propositions of the Kripke model. The \textit{Protocols} of the Agent correspond to the labelling function of Kripke model. The \textit{Protocols} describe which actions can be performed in each state, and that corresponds to which atomic proposition those hold in each state. The \textit{Evolution} functions for an agent describes how the states transit as a result of the actions performed by all other agents, which correspond to the transition relations of the Kripke model that describe the condition of the state transitions. Following this principle, the Kripke models of communication, communication detector traffic information, AIS, fault event, fault detector, the USV and the GCS were translated into the ISPL code.

\subsection{Modelling of properties to be verified}
Considering the real world missions of the C-Enduro USV, there are fourteen properties verified, given as below:

\begin{figure}[!t]
\begin{center}
    \includegraphics[width=2.6in]{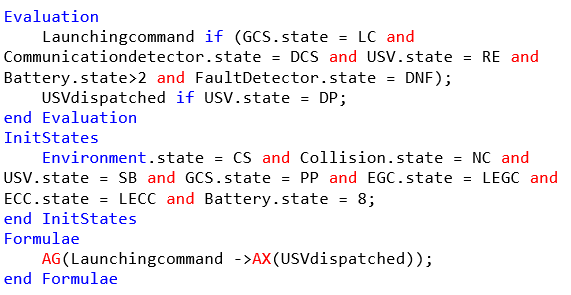}
\caption{ISPL code for Evaluation, Formulae and InitStates}
\label{fevaluationformulae}
\end{center}
\end{figure}

\begin{enumerate}
  \item After the USV received the mission (ready state), if the communication state is good; no fault detected by the USV; and the USV battery level is above 2, then the GCS sends the launching command and the USV will transit to \textit{DP}.

  \item When the USV is in \textit{PF}; no fault is detected; the USV battery level is above 2; and a give-way collision risk is detected, then the USV will always change its way to avoid collision.

  \item If the give-way collision does not appear, the USV will never alter its way to avoid the collision.

   \item After the USV avoided the collision risk, if the communication state is good; no give-way collision risk and fault are detected; and the USV battery level is above 2, then the USV will always continue to follow the path at its normal speed.

   \item When the USV is in \textit{SK}; no fault is detected; GCS is in the situation analysis state under good communication state, the GCS will re-plan the path. Formula 5 is used for checking if the GCS can perform path replanning behaviours successfully when the communication gets recovered.

   \item When the USV is in \textit{PF}, if there is no give-way collision risk and fault event detected; the USV battery level is above 2; and the communication is lost, the USV will change to station keeping state. Formula 6 is for checking the part of safety property that when the USV lost communication signals, it will transit to \textit{SK} until the signal gets recovered.

   \item When the USV is in \textit{SK}, if the communication state is good and the USV battery level is above 2, after the GCS send the new waypoints, the USV will change to path following state.

   \item When the USV is in \textit{PF} and no fault detected, if communication lost is detected, the USV will change to \textit{SK}.

   \item If the communication is not lost or the USV battery level is not below 3 or there is no fault detected, the USV will not station keeping. Formula 9 means the USV will only trigger the \textit{SK} behaviour under the right situations (Communication lost, battery level is low or fault detected).

   \item If communication is lost, the USV will change to \textit{SB} or \textit{SK}. Formula 10 is used for checking the safety property.
   
   \item If the USV is in severe fault state, the USV will change to the \textit{SB} state directly. Formula 11 represents that under the severe fault event situation, the USV will transit to \textit{SB} directly for safety.

   \item If the battery level is less than 2, the USV will not follow the path.

   \item When the USV is in \textit{PF} or \textit{CA}; the battery level is 9; energy generation is high; energy consumption is low; no give-way collision risk detected; and no fault detected, the USV will change to \textit{PFH}.

   \item If the battery level is not above 8, the USV will never transit to \textit{PFH}.

\end{enumerate}

The \textit{CTL} Formula of the first property is given below for demonstration:

\begin{align*}
  AG((USV.state &= RE \wedge Communicationdetector.state\\
  &= DCS \wedge Faultdetector.state \\
  &= DNF \wedge GCS.state \\
  &= LC \wedge Battery.state \\
  &> 2) \rightarrow AX(USV.state = DP))
\end{align*}

Note that $AG$ and $AX$ are CTL operator: $AG(p)$ means along All paths $p$ holds Globally; $AX(p)$ means along All paths, $p$ holds in the neXt state. The verification of Formula 1 and initial states are expressed in \textit{Evaluation}, \textit{Formulae} and \textit{InitStates}, as shown in Fig.~\ref{fevaluationformulae}. Other Formulas were also translated from their \textit{CTL} accordingly.


\subsection{Verification result and analysis}
\begin{figure}[!t]
  \begin{center}
      \includegraphics[width=3.2in]{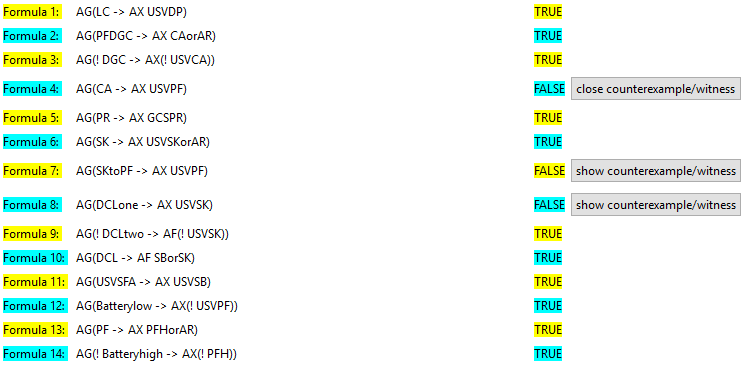}
    \caption{Verification results}
    \label{fresult}
  \end{center}
\end{figure}

The program was executed on a 2.7 GHz Intel Core i7-6820HK processor with 16.0 GB RAM. The number of reachable states approached 209286 when the decision making system was verified and the execution time was 0.632 seconds. The verification results are shown in Fig.~\ref{fresult}. In the verification results, Formula 4, Formula 7 and Formula 8 have FALSE result and the other Formulas have TRUE results. Formula 1 acquires the TRUE result and it shows that the launching behaviour can perform well. The verification results of Formula 2 and 3 show that the collision avoidance command can be executed properly. The verification results of Formula 12, 13 and 14 show that the USV possesses the long endurance/energy saving properties: When the battery level is low, the USV will be in \textit{SB} or \textit{SK}; When the battery level is high, and the energy generation is higher than the energy consumption, the USV will be in \textit{PFH}; If the battery level is not above 8, the USV will never travel in high speed.

\begin{figure}[!t]
\begin{center}
    \includegraphics[width=2in, height=2in]{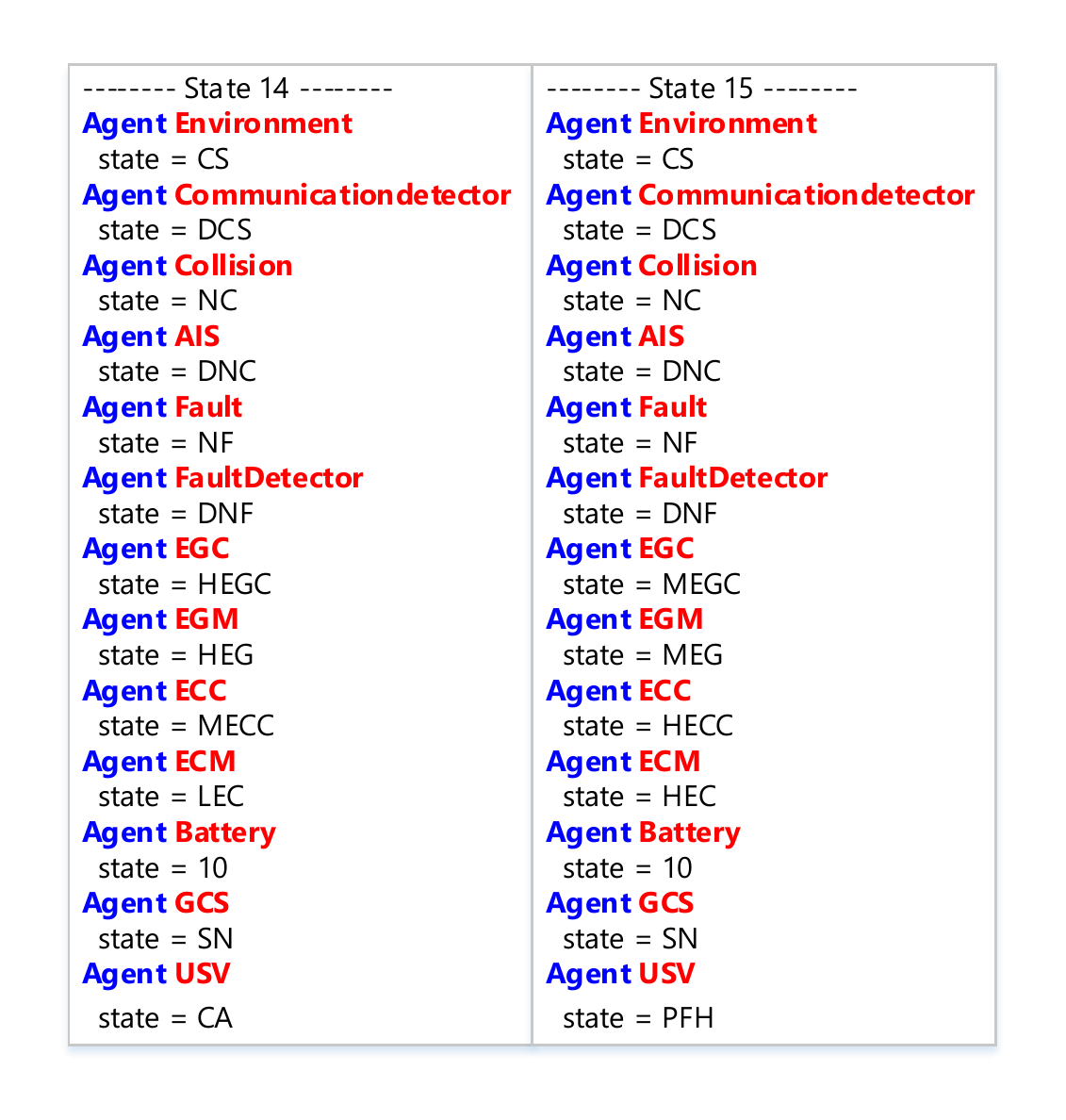}
\caption{Counterexample of Formula 4}
\label{fformula4}
\end{center}
\end{figure}

Using the show counterexample/witness option, the error trace of Formula 4, Formula 7 and Formula 8 can be acquired. By checking the counterexample of Formula 4, as shown in Fig.~\ref{fformula4}, we found the USV transits to \textit{PFH} instead of \textit{PF}, because the record shows that the battery level was 10 and the energy generation was higher than energy consumption. It is reasonable to accelerate to maximise the utilisation of the natural energy. When Formula 4 is changed to ``USV will transit to \textit{PF} or \textit{PFH}'', the verification result became TRUE. The verification record of Formula 7 shows that after the GCS sent a new waypoint and the USV detected a collision risk, so it is transited to collision avoidance state instead of path following to ensure USV safety. Therefore, this counterexample is reasonable. The verification result of Formula 8 shows that when the USV is following the path, and the communication is lost and it also detected the give-way collision risk at the same time, it will transit to collision avoidance state first instead of the station keeping state, compliant with the safety design of the decision making system.

\section{Conclusion and Future Work} \label{sec:conclusion}
This research tackled the problem of applying model checking method for verifying the decision-making behaviours of a long endurance USV, which may encounter communication lost, collision risk, malfunction and maximising energy utilisation problems. The Kripke model and \textit{CTL} were applied to construct the model of environmental factors and autonomous systems. Finally, both the safety properties and the long endurance properties of the decision making behaviours under concurrent and non-deterministic uncertainties were verified using MCMAS. The short program executing time (0.632 seconds) also implies that more complex scenario and more agents can be handled by model checker MCMAS. In the future work, multiple USVs cooperation or UAVs-USVs cooperation can be taken into account in complex scenarios. State space reduction techniques may be required for saving computing resources. A translation programme which transforms a system design to a model checker language will reduce the potential mistakes from the designer.

\bibliographystyle{IEEEtran}
\bibliography{IEEEexample}

\begin{thebibliography}{10}
\providecommand{\url}[1]{#1}
\csname url@samestyle\endcsname
\providecommand{\newblock}{\relax}
\providecommand{\bibinfo}[2]{#2}
\providecommand{\BIBentrySTDinterwordspacing}{\spaceskip=0pt\relax}
\providecommand{\BIBentryALTinterwordstretchfactor}{4}
\providecommand{\BIBentryALTinterwordspacing}{\spaceskip=\fontdimen2\font plus
\BIBentryALTinterwordstretchfactor\fontdimen3\font minus
  \fontdimen4\font\relax}
\providecommand{\BIBforeignlanguage}[2]{{%
\expandafter\ifx\csname l@#1\endcsname\relax
\typeout{** WARNING: IEEEtran.bst: No hyphenation pattern has been}%
\typeout{** loaded for the language `#1'. Using the pattern for}%
\typeout{** the default language instead.}%
\else
\language=\csname l@#1\endcsname
\fi
#2}}
\providecommand{\BIBdecl}{\relax}
\BIBdecl

\bibitem{ren2020finite}
Z.~Ren, B.~Zhao, and D.~T. Nguyen, ``{Finite-Time Backstepping of a Nonlinear
  System in Strict-Feedback Form: Proved by Bernoulli Inequality},'' \emph{IEEE
  Access}, vol.~8, pp. 47\,768--47\,775, 2020.

\bibitem{niu2016efficientplanning}
H.~Niu, Y.~Lu, A.~Savvaris, and A.~Tsourdos, ``{Efficient Path Planning
  Algorithms for Unmanned Surface Vehicle},'' \emph{IFAC-PapersOnLine},
  vol.~49, no.~23, pp. 121--126, 2016.

\bibitem{zhu2020adaptive}
M.~Zhu, W.~Sun, A.~Hahn, Y.~Wen, C.~Xiao, and W.~Tao, ``Adaptive modeling of
  maritime autonomous surface ships with uncertainty using a weighted ls-svr
  robust to outliers,'' \emph{Ocean Engineering}, vol. 200, p. 107053, 2020.

\bibitem{zhu2019optimized}
M.~Zhu, A.~Hahn, Y.-Q. Wen, and W.-Q. Sun, ``Optimized support vector
  regression algorithm-based modeling of ship dynamics,'' \emph{Applied Ocean
  Research}, vol.~90, p. 101842, 2019.

\bibitem{zhu2018identification}
M.~Zhu, A.~Hahn, and Y.-Q. Wen, ``Identification-based controller design using
  cloud model for course-keeping of ships in waves,'' \emph{Engineering
  Applications of Artificial Intelligence}, vol.~75, pp. 22--35, 2018.

\bibitem{bertram2008unmanned}
V.~Bertram, ``Unmanned surface vehicles-a survey,'' \emph{Skibsteknisk Selskab,
  Copenhagen, Denmark}, vol.~1, pp. 1--14, 2008.

\bibitem{niu2019voronoi}
H.~Niu, A.~Savvaris, A.~Tsourdos, and Z.~Ji, ``Voronoi-visibility roadmap-based
  path planning algorithm for unmanned surface vehicles,'' \emph{Journal of
  Navigation}, vol.~72, no.~4, pp. 850--874, 2019.

\bibitem{makhsoos2019evaluation}
A.~Makhsoos, H.~Mousazadeh, and S.~S. Mohtasebi, ``Evaluation of some effective
  parameters on the energy efficiency of on-board photovoltaic array on an
  unmanned surface vehicle,'' \emph{Ships and Offshore Structures}, vol.~14,
  no.~5, pp. 492--500, 2019.

\bibitem{ren2021active}
Z.~Ren, R.~Skjetne, A.~S. Verma, Z.~Jiang, Z.~Gao, and K.~H. Halse, ``Active
  heave compensation of floating wind turbine installation using a catamaran
  construction vessel,'' \emph{Marine Structures}, vol.~75, no. 102868, 2021.

\bibitem{sirigineedi2010kripke}
G.~Sirigineedi, A.~Tsourdos, B.~White, and R.~Zbikowski, ``Kripke modelling and
  model checking of a multiple {UAV} system monitoring road network,'' in
  \emph{Proceedings of the AIAA Guidance, Navigation, and Control Conference},
  vol.~4, 2010.

\bibitem{webster2014generating}
M.~Webster, N.~Cameron, M.~Fisher, and M.~Jump, ``Generating certification
  evidence for autonomous unmanned aircraft using model checking and
  simulation,'' \emph{Journal of Aerospace Information Systems}, vol.~11,
  no.~5, pp. 258--279, 2014.

\bibitem{ezekiel2011verifying}
J.~Ezekiel, A.~Lomuscio, L.~Molnar, S.~M. Veres, and M.~Peabody, ``Verifying
  fault tolerance and self-diagnosability of an autonomous underwater
  vehicle,'' \emph{IJCAI-11 : 22nd International Joint Conference on Artificial
  Intelligence Workshop. AIl in Space: Intelligence beyond Planet Earth,
  Barcelona, Spain. 15 - 21 Jul 2011. 6 pp}, 2011.

\bibitem{molnar2009system}
L.~Molnar and S.~Veres, ``System verification of autonomous underwater vehicles
  by model checking,'' in \emph{OCEANS 2009-EUROPE}.\hskip 1em plus 0.5em minus
  0.4em\relax IEEE, 2009, pp. 1--10.

\bibitem{Clarke:2001:MC:778522.778533}
\BIBentryALTinterwordspacing
E.~M. Clarke and B.-H. Schlingloff, ``{Handbook of Automated Reasoning},''
  A.~Robinson and A.~Voronkov, Eds.\hskip 1em plus 0.5em minus 0.4em\relax
  Amsterdam, The Netherlands, The Netherlands: Elsevier Science Publishers B.
  V., 2001, ch. Model Checking, pp. 1635--1790. [Online]. Available:
  \url{http://dl.acm.org/citation.cfm?id=778522.778533}
\BIBentrySTDinterwordspacing

\bibitem{choi2015verification}
J.~Choi, S.~Kim, and A.~Tsourdos, ``Verification of heterogeneous multi-agent
  system using {MCMAS},'' \emph{International Journal of Systems Science},
  vol.~46, no.~4, pp. 634--651, 2015.

\bibitem{barbier2019validation}
M.~Barbier, A.~Renzaglia, J.~Quilbeuf, L.~Rummelhard, A.~Paigwar, C.~Laugier,
  A.~Legay, J.~Iba{\~n}ez-Guzm{\'a}n, and O.~Simonin, ``Validation of
  perception and decision-making systems for autonomous driving via statistical
  model checking,'' in \emph{2019 IEEE Intelligent Vehicles Symposium
  (IV)}.\hskip 1em plus 0.5em minus 0.4em\relax IEEE, 2019, pp. 252--259.

\bibitem{clarke2018handbook}
E.~M. Clarke, T.~A. Henzinger, H.~Veith, and R.~Bloem, \emph{Handbook of model
  checking}.\hskip 1em plus 0.5em minus 0.4em\relax Springer, 2018, vol.~10.

\bibitem{brat2005challenges}
G.~Brat and A.~Jonsson, ``Challenges in verification and validation of
  autonomous systems for space exploration,'' in \emph{Neural Networks, 2005.
  IJCNN'05. Proceedings. 2005 IEEE International Joint Conference on},
  vol.~5.\hskip 1em plus 0.5em minus 0.4em\relax IEEE, 2005, pp. 2909--2914.

\bibitem{pecheur2000verification}
C.~Pecheur, ``Verification and validation of autonomy software at {NASA},''
  National Aeronautics and Space Administration, Tech. Rep., 08 2000.

\bibitem{quottrup2004multi}
M.~M. Quottrup, T.~Bak, and R.~Zamanabadi, ``Multi-robot planning: {A} timed
  automata approach,'' in \emph{Robotics and Automation, 2004. Proceedings.
  ICRA'04. 2004 IEEE International Conference on}, vol.~5.\hskip 1em plus 0.5em
  minus 0.4em\relax IEEE, 2004, pp. 4417--4422.

\bibitem{sirigineedi2009towards}
G.~Sirigineedi, A.~Tsourdos, B.~A. White, and R.~Zbikowski, ``Towards
  verifiable approach to mission planning for multiple {UAVs},'' in
  \emph{Proceedings of AIAA Infotech@ Aerospace Conference and AIAA Unmanned..
  Unlimited Conference}, 2009.

\bibitem{sirigineedi2011kripke}
G.~Sirigineedi, A.~Tsourdos, B.~A. White, and R.~{\.Z}bikowski, ``Kripke
  modelling and verification of temporal specifications of a multiple {UAV}
  system,'' \emph{Annals of Mathematics and Artificial Intelligence}, vol.~63,
  no.~1, pp. 31--52, 2011.

\bibitem{jeyaraman2006kripke}
S.~Jeyaraman, A.~Tsourdos, R.~{\.Z}bikowski, and B.~White, ``Kripke modelling
  approaches of a multiple robots system with minimalist communication: a
  formal approach of choice,'' \emph{International journal of systems science},
  vol.~37, no.~6, pp. 339--349, 2006.

\bibitem{humphrey2013model}
L.~Humphrey and M.~Patzek, ``Model checking human-automation {UAV} mission
  plans,'' in \emph{Proceedings of the AIAA Guidance, Navigation, and Control
  (GNC) Conference}, 2013.

\bibitem{humphrey2012model}
L.~Humphrey, ``Model checking {UAV} mission plans,'' in \emph{Proceedings of
  AIAA Conference on Modeling and Simulation Technologies}, 2012.

\bibitem{ASV2017}
\BIBentryALTinterwordspacing
{ASV Global}. (2017) {ASV to Build USV under SBRI Funding}. [Online; accessed
  19-June-2020]. [Online]. Available:
  \url{https://www.asvglobal.com/asv-to-build-usv-under-sbri-funding/}
\BIBentrySTDinterwordspacing

\bibitem{huth2004logic}
M.~Huth and M.~Ryan, \emph{{Logic in Computer Science: Modelling and reasoning
  about systems}}.\hskip 1em plus 0.5em minus 0.4em\relax Cambridge university
  press, 2004.

\bibitem{niu2018energy}
H.~Niu, Y.~Lu, A.~Savvaris, and A.~Tsourdos, ``An energy-efficient path
  planning algorithm for unmanned surface vehicles,'' \emph{Ocean Engineering},
  vol. 161, pp. 308--321, 2018.

\bibitem{niu2020energy}
H.~Niu, Z.~Ji, A.~Savvaris, and A.~Tsourdos, ``Energy efficient path planning
  for unmanned surface vehicle in spatially-temporally variant environment,''
  \emph{Ocean Engineering}, vol. 196, p. 106766, 2020.

\bibitem{savvaris2014development}
A.~Savvaris, H.~Niu, H.~Oh, and A.~Tsourdos, ``Development of collision
  avoidance algorithms for the c-enduro usv,'' in \emph{Proceedings of the 19th
  IFAC world congress (IFAC 2014)}, vol.~47, 2014, pp. 12\,174--12\,181.

\end{thebibliography}

\end{document}